\documentclass[a4paper,10pt]{article}

\usepackage{geometry}
\usepackage[utf8]{inputenc}
\usepackage[T1]{fontenc}
\usepackage{ae,aecompl}

\usepackage{enumitem}

\usepackage[authoryear,round]{natbib}
\usepackage{color}
\usepackage[colorlinks=true,linkcolor=blue,citecolor=red]{hyperref}

\usepackage{amsfonts,amssymb,mathtools,amsthm}

\newtheorem{theorem}{Theorem}

\theoremstyle{definition}
\newtheorem{definition}{Definition}

\theoremstyle{remark}
\newtheorem{remark}{Remark}

\newcommand{\Rb}{\mathbb R}
\newcommand{\Nb}{\mathbb N}

\newcommand\Cc{\mathcal C}

\newcommand{\veps}{\varepsilon}
\newcommand{\vphi}{\varphi}

\newcommand{\indic}[1]{\mathbf{1}_{#1}}

\newcommand{\ds}{\displaystyle}
\newcommand{\viz}{\textit{viz.}}
\newcommand{\eg}{\textit{e.g.}}

\newcommand{\cin}{c^{\mathrm{in}}}

\title{Deterministic and Stochastic Becker-D\"oring equations: \\ Past and Recent Mathematical Developments}
\author{E.~Hingant \and R.~Yvinec}
\date{\today}

\newcommand{\Addresses}{{
  \bigskip
  \footnotesize

  E.~Hingant, \textit{Departmento de Matem\'atica, Universidad del Bío-Bío,
    Concepci\'on, Chile}\par\nopagebreak
  \textit{E-mail address}: \texttt{ehingant@ubiobio.cl}

  \medskip

  R.~Yvinec, \textit{Physiologie de la Reproduction et des Comportements, Institut 
National de la Recherche Agronomique (INRA) UMR85, CNRS-Universit\'e Fran\c{c}ois-Rabelais UMR7247, IFCE, Nouzilly, 
37380 France}\par\nopagebreak
  \textit{E-mail address} \texttt{romain.yvinec@tours.inra.fr}

}}

\begin{document}

\maketitle

\setcounter{tocdepth}{2}
\tableofcontents


\section{Introduction}

The Becker-D\"oring (BD) equations goes back to the seminal work \emph{``Kinetic treatment of nucleation in supersaturated vapors''}~by \cite{Becker1935}, which gave rise to the name of the model. Later on, \cite{Burton1977} popularized the use of such equations to study condensations phenomena at different pressures. Since then, applications of this model range from physics, chemistry to biology. Recently, the book edited by \cite{Schmelzer2005} make an inventory of several applications of nucleation and phase transition theory. Let us also point out recent applications of Becker-D\"oring or related coagulation-fragmentation models in biology, specifically to protein aggregation in neurodegenerative diseases \textit{e.g.}~the works by \cite{Linse2011}, \cite{Prigent2012}, \cite{Alvarez-Martinez2011}, \cite{Budrikis2014}, \cite{Eden2015}, \cite{Davis2016}, \cite{Eugene2016} and \cite{Doumic2016}, also \cite{Hu2011} worked on polymerization of actin filaments, \cite{Hoze2014b,Hoze2014} on assembly of virus capsids, \cite{Bressloff2016} on vesicular transport, and \cite{Hoze2012} for telomere clustering. 

 In its survey, \cite{Slemrod2000} said the BD equations {\it ``provide perhaps the simplest kinetic model to describe a number of issues in the dynamics of phase transitions''}. This is maybe one of the reason these equations received lot of attentions from many mathematicians. But being simple these equations do not prevent its richness and difficulties. Our intention here, is: on one hand, to complete the review by Slemrod with new results; and on the other hand, to give a parallel with the stochastic version of these equations, which reveals a lot of new interesting problems. We also mention the review by \cite{Wattis2006} which contains many qualitative and exact properties of the solutions in the deterministic context, and the pedagogical notes by \cite{penrose2001}. But, few stochastic review of the BD model is available, we can only mention the seminal work by \cite{Aldous1999} which treats the so-called Smolukowsky coagulation equations. 

The model consists in describing the repartition of clusters by their size $i\geq 1$, {\it i.e.} the number of particles that composed them. Clusters belong to a ``solvent'' in much smaller proportion and are assumed to be spatially homogeneously distributed. Along their motion, clusters give rise to two types of reactions, namely the \textit{Becker-D\"oring rules}:

\begin{enumerate}
 \item A cluster of size $1$, commonly called {\it monomer} or {\it elementary particle}, may encounter a cluster of size $i\geq 1$ to coalesce and give rise to a cluster of size $i+1$.
 \item A cluster of size $i\geq 2$ may release spontaneously a monomer resulting in a cluster of size $i-1$. 
\end{enumerate}
These can be summarized by the set of kinetic reactions, for each $i\geq 1$, 
\begin{equation}\label{eq:kinetic_reaction}
   C_1 + C_i  \xrightleftharpoons[b_{i+1}]{a_i}  C_{i+1} \,,
\end{equation}
where $C_i$ denotes clusters consisting of $i$ particles. Coefficients $a_i$ and $b_{i+1}$ stand, respectively, for the rate of aggregation and fragmentation. These may depend on the size of clusters involved in the reactions and typical coefficients are derived by \cite{Penrose1997} and \cite{Niethammer2003}:
\begin{equation}\label{eq:example_coefficients}
 a_i=i^\alpha\,,\quad b_{i+1}=a_{i+1}\left(z_s+\frac{q}{(i+1)^{\gamma}}\right)\,, \qquad i\geq 1\,.
\end{equation}
for $0\leq \alpha < 1$, $z_s>0$, $q>0$ and $0<\gamma<1$. This choice is in agreement with original derivation where $a_i \approx i^{2/3}$, $b_i\approx a_i \exp(G i^{-1/3})$. In particular, the diffusion-limited case of monomers clustering into sphere corresponds to  $\alpha=1/3$, $\gamma=1/3$ in 3D and to $\alpha=0$, $\gamma=1/2$ in 2D, while the interface-reaction-limited case corresponds to  $\alpha=2/3$, $\gamma=1/3$ in 3D and $\alpha=1/2$, $\gamma=1/2$ in 2D.
%
We refer also to \cite{Penrose1983} for a method on deriving coefficients. Note that all along the survey we assume the natural hypothesis $a_i$ and $b_{i+1}$ are non-negative for all $i\geq1$, without referring to this again.  

In  its mean-field version, or deterministic, the BD model is an infinite set of ordinary differential equations for the time evolution of each concentrations (numbers per unit of volume) of clusters made of $i$ particles. In its stochastic version, the BD model is a continuous time Markov chain, on a finite state space. We divide the remainder of this survey in two parts for the respective versions.

\section{Deterministic mean-field theory}

The general formulation of the deterministic Becker-D\"oring equations, as studied today, seems to go back to \cite{Burton1977} and was popularized among mathematicians by \cite{Penrose1979} (indeed, the equations studied in the original work by \cite{Becker1935} slightly differ, see comment later on). It assumes the system behaves homogeneously in space with a high number of clusters, and considers concentrations $c_i(t)$ (unit per volume) of clusters with size $i\geq 1$ at time $t\geq0$. It deals with classical law of chemistry (Law of Mass Action), the coagulation is considered as a second order reaction  while the fragmentation is a first-order (linear) reaction. The flux associated to the kinetic scheme \eqref{eq:kinetic_reaction} is thus given, for each $i\geq 1$, by  

\begin{equation}\label{eq:BD_flux}
J_{i} = a_i c_1 c_i - b_{i+1} c_{i+1}\,.
\end{equation}
Considering all the fluxes involved in the variations of the concentration of each $c_i$ entails the infinite system of differential equations, namely the Becker-D\"oring equations:
\begin{align}
 &\frac{d}{dt} c_1  =   -  J_{1} - \sum_{i \geq {1}} J_i \,, \label{sys:BD-1}\\
 & \frac{d}{dt} c_{i}  = J_{i-1} -  J_i \,, \label{sys:BD-2}
\end{align}
for every $i\geq 2$. The system considered here has no source nor sink. Consequently, for the total amount of monomers, we should have, for all $t\geq 0$,
\begin{equation}\label{eq:BD_mass}
\sum_{i\geq 1}  i c_i(t) =  \rho \,,
\end{equation}
where $\rho$ is a constant, called through the survey: \textit{mass of the system}. Formal computations on the solution of the system (\ref{sys:BD-1}-\ref{sys:BD-2}), interverting infinite sum, lead to this statement. Remark, the constant $\rho$ is entirely determined by the  initial condition given at time $t=0$. In this section we try to expound the main theory  around these equations. In particular, we exclude many variants such as the original constant monomer formulation, which is then an infinite linear system (\eg~\citealp{Penrose1989}, \citealp{Kreer1993} or \citealp{King2002}), the finite-dimensional truncated system (\eg~\citealp{Duncan2001} or \citealp{Duncan2002}), generalization such as micelles formation \citep[\eg][]{Coveney} or including space with cluster diffusion \citep[\eg][]{Laurencot1998} or lattice models \citep[\eg][]{Penrose1983}.

We separated this section between  well-posedness, long-time behavior, scaling limit, and some time-dependent properties.

\subsection{Well-posedness}

The first \textit{general} result on existence and uniqueness on Becker-D\"oring equations is due to \cite{Ball1986} which really start the mathematical analysis of BD equations. The authors state many of the fundamental properties of the solutions belonging to the Banach space
\[X^+ := \left\{ x \subset \Rb_+^\Nb \, : \, \sum_{i\geq 1} i x_i < + \infty \right\}\,,\]
which arises naturally in view of the balance of mass Eq. \eqref{eq:BD_mass}. We recall first the notion of solutions to BD equations.
\begin{definition}
 Let $T\in(0,+\infty]$ and $\cin \in X^+$. A solution to the Becker-D\"oring equations~(\ref{sys:BD-1}-\ref{sys:BD-2}) on $[0,T)$ with initial data $c^{\rm in}$, is a function $c:[0,T) \to X^+$  which writes $c:=(c_i)_{i\geq 1}$ and such that: $\sup_{t\in[0,T)} \|c(t)\|_X <+\infty$; for all $t\in[0,T)$, we have \smash{$\sum_{i\geq 1} a_i c_i \in L^1(0,t)$} and \smash{$\sum_{i\geq 2} b_i c_i \in L^1(0,t)$}; Eqs.~(\ref{sys:BD-1}-\ref{sys:BD-2}) hold almost every $t\in[0,T)$ and $c(0)=\cin$.
\end{definition}
One of the fundamental fact, proved by \cite{Ball1986}, is that any solution to the BD equations satisfies the balance of mass Eq. \eqref{eq:BD_mass} at all finite time (Corolary 2.6). In particular, any solution to the BD equations avoids the so-called gelation phenomenon (in finite time) which can occur in general coagulation-fragmentation equations (\eg~\citealt{Escobedo2003}). \cite{Ball1986} also proved propagation of moments (Theorem 2.2) and regularity properties of the solutions (Theorem 3.2). Finally, they state a general existence result for sublinear coagulation rate and uniqueness with an extra-moment on the initial condition (see below Theorem \ref{thm:BD_wellposed}). In short, the work by \cite{Ball1986} covered the essential properties of BD equations, build the foundations for the analysis of BD equations, and should be a companion for whom want to work with.    

We go back to well-posedness, for which \cite{Laurencot2002a} complement the result, by  \cite{Ball1986}, proving the uniqueness without extra condition on the initial data but assuming a growth condition on the fragmentation rate, \viz~there exists a constant $K>0$ such that, for every $i\geq 2$, 
\begin{equation} \label{eq:growth_coefficient}
 a_{i} - a_{i-1} \leq K \,,  \quad  b_{i} - b_{i+1} \leq K\,.
\end{equation}
We summarize these results in the following theorem.
\begin{theorem}[Well-posedness,  \cite{Ball1986}, \cite{Laurencot2002a}] \label{thm:BD_wellposed}
Let $\cin \in X^+$. Assume alternatively either a) $a_i  = O(i)$ and $\sum_{i\geq 1} i^2 c_i^{\rm in} < +\infty$, or b) the growth condition \eqref{eq:growth_coefficient}. The Becker-D\"oring equations~\textup{(\ref{sys:BD-1}-\ref{sys:BD-2})} have a unique solution $c$ on $[0,+\infty)$ associated with the initial data $\cin$. Moreover, for all $t\geq 0$,
\begin{equation*}
   \sum_{i\geq 1} i c_i(t) = \sum_{i\geq 1} i \cin_i\,.
\end{equation*}
\end{theorem}

In fact the uniqueness by \cite{Ball1986} is slightly more subtle, see their Theorem 3.6. Also, they proved that $a_i=O(i)$ is almost optimal. Indeed, their Theorem 2.7 states: if $\lim_{i\to\infty} a_i/i = +\infty$ and $\lim_{i\to \infty} b_{i+1}/{a_i} < +\infty$, then for some initial condition (with relatively {\it fat} tail) still belonging to $X^+$ the BD system has no-solution. This suggests that, for super-linear coagulation rate, we cannot hope existence for a large class of initial data without a sufficient control on the fragmentation rate. Since mass is preserved,  fragmentation should balance the formation of ``big'' clusters. It seems very few results exist for such class of coefficients, except \cite{Wattis2004a} who considered exponential coefficients.

Finally, we mention that a proof of existence to the BD equations is self-contained in the nice proof by \cite{Laurencot2002} for a more general model (discrete coagulation with multiple fragmentation). It relies, as for the proof given by \cite{Ball1986}, on a truncated system up to a size $N$ and compactness arguments to obtain the limit $N\to+\infty$. But here \cite{Laurencot2002} took advantage of the propagation of super-linear moments and a De La Vallé Poussin lemma to prove compactness.   

\subsection{Long-time behavior}

The long-time behavior of the BD system brings some of its most interesting properties, and we will see this is still under active research. Through this section we will always assume that both $a_i$ and $b_{i+1}$ are positive for each $i\geq 1$. This avoids many \textit{pathological} cases, in some sense, if one of them cancel it ``breaks the communication'' between clusters in one side or another. Nonetheless, we mention the interesting cases (not detailed here) where either $a_i=0$ or $b_{i+1}=0$, for every $i\geq 1$, which have been treated again by \cite{Ball1986}! We start with a subsection which deals with convergence to equilibrium. Then, we will see the most recent results on the exponential stability of the equilibrium.

\subsubsection{Convergence to equilibrium}

The equilibrium candidates, at plural, of the BD equations are obtained by canceling the fluxes $J_i$, for each $i\geq 1$, as defined in Eq.~\eqref{eq:BD_flux}. After straightforward manipulation of the fluxes, the candidates form a one-parameter family, indexed by a variable $z\geq 0$, and are given by the expressions
\begin{equation}\label{eq:equilibrium}
 \bar c_i(z) = Q_i z^i\,, \quad \mathrm{ where } \quad   Q_i = \frac{a_1a_2\cdots a_{i-1}}{b_2b_3\cdots b_i}\,,
\end{equation}
for every $i\geq 1$, with the convention $Q_1=1$. For example, the case related to Eq.~\eqref{eq:example_coefficients} gives \citep[\eg][]{Niethammer2003}, for large $i$,
\begin{equation*}
Q_i \approx \frac{C}{i^{\alpha}z_s^{(i-1)}}\exp\left(-\frac{q}{(1-\gamma)z_s}i^{1-\gamma}(1+O(i^{-\gamma}))\right)\,.
\end{equation*}
To find the right equilibrium, which reduces to find the value of $\bar c_1=z$, one should use the balance of mass Eq.~\eqref{eq:BD_mass} which we know to be satisfied at any finite time. Hence, this leads us to consider the power series, given by the mass of the equilibrium candidates, 
\begin{equation*}
 \sum_{i\geq 1} iQ_i z^i \,,
\end{equation*}
which radius of convergence is denoted by $z_s$. Such radius is obtained from the rates functions since the Cauchy-Hadamard theorem says \smash{$1/z_s := \limsup_{i\to\infty} Q_i^{1/i}$}. This becomes the heart of the existence of a critical mass in BD equations since the values taken by the series may not define a bijection from $[0,z_s)$ into $[0,+\infty)$. Indeed, set $\rho_s$ be the upper value taken by the series on $\{z< z_s\}$. It can occur that $\rho_s$ is finite, in which case we already know that there is no equilibrium candidate with mass $\rho>\rho_s$. Hence, this leads to a dichotomy in the long-time behavior of the BD equations whether or not the mass of the solution considered is less than $\rho_s$, named the \textit{critical mass}. We may refer to sub-critical solution when the mass $\rho < \rho_s$, critical solution when $\rho=\rho_s$ and super-critical solution when $\rho>\rho_s$.

The Becker-D\"oring equations are part of the kinetic equations. These later have a long story, led by the celebrated Boltzmann equations, which are of course completely out of the scope of this paper, maybe the reader could refer to \cite{Cercignani1990}. The key concept in these equations is the \textit{entropy} (sometimes called energy) which, in  mathematical words, is a Lyapounov functional and governs the trend to equilibrium. Namely, the entropy arising in BD equations is given by the expression
\begin{equation*}
H(c) = \sum_{i\geq 1} c_i \left( \ln\left(\frac{c_i}{Q_i}\right)-1\right)\,.
\end{equation*}
This is because, formally, the $H$ decreases along the solutions $c$ (and is bounded from below), as
\begin{equation} \label{eq:H-theorem}
\frac{d}{dt} H(c(t)) = -D(c(t))\,,
\end{equation}
where the \textit{dissipation} is
\begin{equation*}
D(c):=-\sum_{i\geq 1} \left(a_ic_1c_i-b_{i+1}c_{i+1} \right)\left(\ln (a_ic_1c_i)-\ln(b_{i+1}c_{i+1})\right)\,.
\end{equation*}
Remark, since $\ln$ is increasing, the dissipation $D$ is non negative.  Depending on the properties you are looking for, this is possible to define the {\it relative entropy} functional, with same dissipation term, and given by the expression
\begin{equation*}
H(c|c^\rho) = \sum_{i\geq 1} c_i \left( \ln\left(\frac{c_i}{c^\rho_i}\right)-1\right) + \sum_{i\geq 1} c^\rho_i \,,
\end{equation*}
where $c^\rho$ is the equilibrium candidate, with mass $\rho$, \textit{i.e.}~the components are given by Eq.  \eqref{eq:equilibrium} for which $z$ is chosen such that \smash{$\sum_{i\geq 1} iQ_iz^i = \rho$}. The second term in the right-hand side, ensuring non-negativity, is some times omitted. Hence, in the case initially $H(\cin|c^\rho)<+\infty$, we should have $D(c(t)) \to 0$ as $t\to \infty$ as we can see in 
\begin{equation} \label{eq:weak-H-theorem}
0 \leq  H(c(t)|c^\rho) + \int_0^t D(c(s))\, ds \leq H(\cin|c^\rho)\,.
\end{equation}
And remarking that $D = 0$ corresponds, see its definition, to $J_i=0$ for all $i\geq 1$, we have a good reason to go ahead with the functional $H$. The hard work is to prove rigorous properties on the entropy and relative entropy, along the solutions. Again \cite{Ball1986} set the basic. The authors give many results, among others, continuity properties of the entropy functional (Proposition 4.5) and minimizing sequence properties (Theorem 4.4). Also, they proved the key ingredient that Eq.~\eqref{eq:weak-H-theorem} holds for a large class of rate (Theorem 4.8). Finally, in their Theorem 4.7, they proved the so-called $H$-theorem (by analogy with the celebrating Boltzmann $H$-theorem), which is a rigorous justification of \eqref{eq:H-theorem}.
\begin{theorem}[H-theorem, \cite{Ball1986}]\label{thm:H-theorem}
 Assume $z_s>0$, $\liminf_{i\to\infty} Q_i^{1/i}>0$, $a_i=O(i/\ln i)$ and $b_i=O(i/\ln i)$. If $c$ is a solution to the Becker-D\"oring equations~\textup{(\ref{sys:BD-1}-\ref{sys:BD-2})} on $[0,T)$, for some $T\in(0,+\infty]$, with initial condition $\cin\neq 0$ belonging to $X^+$, then dissipation of entropy Eq.~\eqref{eq:H-theorem} holds almost every $t\in[0,T)$.
\end{theorem}
Note that linear growth $a_i,\, b_i \sim i$ is not allowed. Fewer assumptions on the rate of fragmentation is possible, adapting the results obtained for general coagulation-fragmentation equation by \cite{Carr1994} and later by \cite{Canizo2007}. Now we state the main asymptotic results. A very general result in the case $z_s=+\infty$ is available from Theorem 5.4 by \cite{Ball1986}. But the more interesting case is $0<z_s<+\infty$ for which a dichotomy  occurs. This is treated for particular initial conditions and rates by \cite{Ball1986}, and then extended to general initial conditions in \cite{Ball1988}. Finally, it was refined by \cite{Slemrod1989} for a class of rates allowing linear growth, see its Theorem 5.11, which we state below. 
\begin{theorem}[Convergence to equilibrium, \cite{Slemrod1989}] \label{thm:convergence_equilibrium}
Let $\cin\in X^+$ with mass $\rho$ and such that $H(\cin)<+\infty$. Assume $a_i = O(i)$, $b_i = O(i)$ and \smash{$\lim_{i\to+\infty} Q_i^{1/i}=1/z_s$} exists $(z_s>0)$. Assume moreover  there exists $z\in[0,z_s]$ such that $a_i z \leq b_i$ for sufficiently large $i$. Finally,  let  $c$ be the unique solution to the Becker-D\"oring equations~\textup{(\ref{sys:BD-1}-\ref{sys:BD-2})} on $[0,+\infty)$ with initial data $c^{\rm in}$. We have: 
\begin{enumerate}[label=\textup{(\alph*)}]
 \item If $0\leq \rho \leq \rho_s$, then $\ds \lim_{t\to+\infty} \, \sum_{i\geq 1} i|c_i(t) - c^{\rho}_i| = 0\,.$
 \item If $\rho > \rho_s$, then, for every $i\geq 1$, $\ds \lim_{t\to+\infty}\, c_i(t) =  c^{\rho_s}_i\,.$
\end{enumerate}
In both case we recall that $c^\rho$ is the equilibrium given by Eq.~\eqref{eq:equilibrium} with mass $\rho$.
\end{theorem}

Surprisingly, in point (b), while the solution has mass $\rho$ for all times, as the time goes to infinity, it converges in a weak sense (component by component) to a solution having a strictly inferior mass. In this theory, the difference $\rho-\rho_s$ is interpreted as the formation of particles with \textit{infinite} size and of different nature, phenomenon called \textit{phase transition}. In Section \ref{ssec:large-clusters} we will describe in more details this phenomenon.

The proof consists first in proving that the $w(\cin)-limit$ set consist of equilibrium candidate $c^{\rho'}$ with mass $\rho'$ less than $\min(\rho,\rho_s)$. This is achieved by compactness of the orbit, analyzing the time-translation, and by regularity of the $c_1$ which requires in particular $b_i=O(i)$ (see Theorem 3.2 in \cite{Ball1986}), contrary to the known existence result stated above in Theorem \ref{thm:BD_wellposed} . Then, the limit is selected  thanks to the dissipation \eqref{eq:weak-H-theorem}. A key ingredient is the continuity property of $c \mapsto H(c|c^\rho)$ which holds if and only if $\lim_{i\to+\infty} Q_i^{1/i}$ exists and $\rho=\rho_s$, see Proposition 4.5 by \cite{Ball1986}. Note, the condition $a_i z \leq b_i$ comes from the Theorem 2 by \cite{Ball1988}, and re-used by \cite{Slemrod1989}, which ensures the tail of the $c_i$'s decays sufficiently fast (fragmentation dominates). We point out that these two last conditions are needed to select the right equilibrium, while convergence to some equilibrium is ``always'' satisfied, see Theorem 5.10 by \cite{Slemrod1989}.

We finish by a comment on the case where $z_s=0$, corresponding to a strong coagulation rate, relatively to the fragmentation. \cite{Carr1999} proved under reasonable assumptions that for all $i\geq 1$,  $c_i(t)\to 0$ as $t\to+\infty$.

\subsubsection{Rate of convergence} 

The natural question that arises after the convergence to equilibrium is the \textit{rate} of convergence. When the $H$-Theorem~\ref{thm:H-theorem} holds, with the relative-entropy for instance, we could hope that convergence holds in this sense. The best situation would be the dissipation bounded from below by the entropy itself, \textit{i.e.}~along the solutions: $D(c(t))\geq C H(c(t)|c^\rho)$ for some constant $C>0$. This leads immediately to an exponential decay of the entropy. Unfortunately this does not hold in all cases. A recent proof for $a_i\sim i$ is given by \cite{Canizo2015}. Another way is to bound from below the dissipation by a non-negative function $\psi$ depending on $H$, leading to 
\[ \frac{d}{dt} H(c(t)|c^\rho)  \leq - \psi(H(c(t)|c^\rho))\,.\]
And the problem resumes to find sub-solutions to this ordinary differential equation. This method is named \emph{entropy entropy-dissipation}, because dissipation is created by entropy itself. But this method does not in general lead to exponential decay of the entropy. The first result in this direction is due to \cite{Jabin2003}. Let us show their result.

\begin{theorem}[Rate of convergence, \cite{Jabin2003}] \label{thm:jabin}
Assume $1 \leq a_i = O(i)$, $1\leq b_i = O(i)$, \smash{$\lim_{i\to+\infty} Q_i^{1/i}=1/z_s$} exists $(z_s>0)$ and that $a_i z_s \leq \min(b_i,b_{i+1})$ for every $i\geq 1$. Suppose moreover that $\cin\in X^+$ with mass $\rho<\rho_s$ (sub-critical case), with $H(\cin|c^\rho)<+\infty$ and there exists $\nu>0$ such that \smash{$\sum_{i\geq 1} \exp(\nu i) \cin_i <+\infty$}.  The solution $c$ to the Becker-D\"oring equations~\textup{(\ref{sys:BD-1}-\ref{sys:BD-2})} on $[0,+\infty)$ with initial data $\cin$ satisfies, for some constant $k$ depending on $\cin$ and for all $t\geq 0$
\begin{equation*}
 H(c(t)|c^\rho) \leq H(\cin|c^\rho) \exp(- k t^{1/3} )\,.
\end{equation*} 
\end{theorem}
This theorem is obtain thanks to $\psi(H) = H/\ln(H)^2$. And the authors were able to go back from this estimate to the convergence in $\exp(-kt^{1/3})$ in the strong norm of $X^+$. Similar results are obtained by \cite{Canizo2015} in various cases allowing fewer hypotheses. But these results still not provide satisfactory rate of decay, with pure exponential decay. A well know theory is the stability of linear operator. If the linearized system is locally exponentially stable, we could hope that so is the full non-linear system, in a small neighborhood of the equilibrium. And we could imagine that this small neighborhood is an absorbing set, since we have Theorem~\ref{thm:jabin}. In fact these steps were followed by \cite{Canizo2013} to obtain their nice proof of the full exponential convergence stated below. 

\begin{theorem}[Exponential stability, \cite{Canizo2013}]
 Under the hypothesis of Theorem \ref{thm:jabin} and in addition  $\lim_{i\to +\infty} a_{i+1}/a_i = z_s \cdot \lim_{i\to \infty} Q_{i+1}/Q_i = 1$. The solution $c$ to the Becker-D\"oring equations~\textup{(\ref{sys:BD-1}-\ref{sys:BD-2})} on $[0,+\infty)$ with initial data $\cin$ satisfies, for all $t\geq 0$
\begin{equation*}
 \| c(t) - c^\rho \| \leq \exp(- \lambda t )\,.
\end{equation*} 
\end{theorem}

The constant $\lambda$ is completely calculable from the constant of the problem, important fact for applicability. We refer to their article \citep{Canizo2013} for a very well-detailed introduction and presentation to the result. We mention that the linear Becker-D\"oring system (with constant monomer $c_1$) also exhibits exponential decay \cite{Kreer1993}, and a quantitative comparison of the convergent rates will be of interest. 

We finish this section  pointing  out that the critical case is still completely open. The super-critical too. But in this case it might not happen for the reason we will detail in the Section \ref{sec:metastability}.

\subsection{Coarsening and relation to transport equation}

From the Becker-D\"oring equations~(\ref{sys:BD-1}-\ref{sys:BD-2}), the reader familiar with numerical analysis may recognize that equations on $c_i$, for every $i\geq 2$, has the flavor of a discretization of a transport equation. To make the link more apparent, it is useful to write down the weak form of Eq.~(\ref{sys:BD-1}-\ref{sys:BD-2}), which is also a very useful tool for the study of the BD system it-self. Take $(\vphi_i)_{i\geq 2}$ a sufficiently regular sequence, we then obtain
\begin{equation}\label{eq:bd_weakform}
\frac{d}{dt}\sum_{i\geq 2} c_i(t)\vphi_i =  \vphi_2 J_1 + \sum_{i\geq 2} \left(\vphi_{i+1}-\vphi_i\right)J_i\,.
\end{equation}
where we recall the fluxes $J_i$ are defined by Eq.~\eqref{eq:BD_flux}. Clearly, $\left(\vphi_{i+1}-\vphi_i\right)$ can be seen as a discrete ``spatial'' derivative. Moreover, assuming some ``spatial continuity'', it is tempting to rewrite $J_i$ as $J_i \approx \left(a_ic_1-b_{i}\right)c_i$. With such ansatz, the last equation \eqref{eq:bd_weakform} then motivates the introduction of the following continuous transport equation (in a weak form)
\begin{equation}\label{eq:ls_weakform}
\frac{d}{dt}\int_0^\infty \vphi(x) f(t,x)dx  =  \vphi(0) N(t) + \int_{0}^\infty \vphi'(x) j(t,x) f(t,x) dx\,,
\end{equation}
where the flux now reads 
\begin{equation*}
j(t,x)=a(x)u(t)-b(x)\,,
\end{equation*}
for some appropriate functions $a$ and $b$, and a function $u$ that plays the role of $c_1$. We will see later what should be $N$ and what becomes the mass conservation stating $u$ in the subsequent sections. Both are the main difficulties of the problem in linking the discrete Eq.~\eqref{eq:bd_weakform} to the continuous Eq.~\eqref{eq:ls_weakform}.

As a matter of fact, they depend crucially on the scaling hypothesis (a small parameter which allows passing from discrete size $i$ to continuous size $x$) and on the kinetic coefficients $a$ and $b$. We note that Eq.~\eqref{eq:ls_weakform} is the weak form of a nonlinear transport equation known as the Lifshitz-Slyozov (LS) equation, after the work by \cite{Lifshitz1961}, 
\begin{equation}\label{eq:LS}
\frac{\partial}{\partial t} f + \frac{\partial}{\partial x} \big(j(t,x)f(t,x)\big) =0\,,
\end{equation}
together with (if appropriate) the boundary condition at $x=0$, 
\begin{equation}\label{eq:BC_LS}
\lim_{x\to 0^+} j(t,x)f(t,x)= N(t)\,,
\end{equation}
and an equation for $u$. Rigorous results making connection from the Becker-D\"oring  Eq.~(\ref{sys:BD-1}-\ref{sys:BD-2}) to the Lifshitz-Slyozov Eq.~\eqref{eq:LS} are of two kinds.
First, in the works initiated by \cite{Laurencot2002a}, \cite{Collet2002}, and pursued in \cite{Deschamps2016}, the authors proved that a suitable rescaling of the solution to BD equations (with the essential assumption of large excess of monomers $c_1$) converges to a solution of LS, on any finite time period and either in density or measure functional spaces. Second, in the works initiated by \cite{Penrose1978} and pursued by \cite{Penrose1997}, \cite{Niethammer2003} and \cite{Niethammer2004}, the authors show that long-time behavior of super-critical  solutions to BD equations are closed to the solution of LS.

\subsubsection{Evolution of large clusters in the super-critical case} \label{ssec:large-clusters}

We saw in Theorem \ref{thm:convergence_equilibrium}, in the case $\rho>\rho_s$, that the solution behaves particularly, as infinitly large clusters are created as time goes to infinity. The idea by \cite{Penrose1997} is to perform a time/space scaling to approach the cluster distribution, both in a very long time and for very large sizes, in order to explain the loss of mass $\rho-\rho_s>0$ in the super-critical case. The formal arguments for coefficients given by Eq.~\eqref{eq:example_coefficients} with $\gamma=\alpha=1/3$, are derived by \cite{Penrose1997}, we refer also to the review by \cite{Slemrod2000}. We present here the rigorous result obtained by \cite{Niethammer2003}, for any coefficients given by Eq.~\eqref{eq:example_coefficients} where the author proved that large clusters obey a variant of LS, named the Lifshitz-Slyozov-Wagner (LSW) equations, see below. For a review on LSW and Ostwald Ripening (out of the scope of this paper), see \cite{Niethammer2006} and \cite{Niethammer2008}.

We sketch the formal arguments following \cite{Penrose1997} and \cite{Niethammer2003}. To consider the behavior of large clusters at large time, we introduce an \textit{ad hoc} small parameter $0<\veps \ll 1$ such that $1/\veps$ will be a measure of a typical large cluster.  Within the particular choice of coefficients given by Eq.~\eqref{eq:example_coefficients}, it turns that a new time scale given by $\tau=\veps^{1-\alpha+\gamma} t$, where $\alpha$ and $\gamma$ are the exponents arising in the coefficients, is an appropriate time scaling to obtain a non trivial dynamics. Indeed, we obtain by the BD equations (\ref{sys:BD-1}-\ref{sys:BD-2}) the reformulation
\begin{equation*}
\frac{d}{d\tau} c_i=\frac{1}{\veps^{1-\alpha+\gamma}}\left(J_{i-1}-J_i\right)\,,
\end{equation*}
and the fluxes $J_i$ in Eq.~\eqref{eq:BD_flux} become 
\begin{equation*}
J_i=a_i\left(c_1-z_s-\frac{q}{i^\gamma}\right)c_i-\left(b_{i+1}c_{i+1}-b_ic_i\right) \,,
\end{equation*}
for every $i\geq 1$. Since large clusters are formed as time goes to infinity, it is possible to consider the system after a (possibly long) time $t_\veps$ for which the relative entropy $H(c|c^{\rho_s})$ is small enough, namely of order $\veps^{\gamma}$. This suggests that the small clusters, up to some cut-off $i_\veps$ are close to their equilibrium value, for $t\geq t_\veps$ and $i\leq i_\veps$, given by

\[ c_i(t) = Q_i z_s^i (1+o(1))\,.\]

On the other hand large cluster may be described by a continuous variable $x=\veps i$ for $i\geq i_\veps$. Thus, we define a density $f$ (stepwise) according to the variable $x\geq \veps i_\veps$ by
\begin{equation*}
 f^\veps(\tau,\veps i) = \frac 1 {\veps^2} c_i(\tau)\,. 
\end{equation*}
Respectively we let $u^\veps(\tau) = (c_1(t)-z_s)/\veps^\gamma$. This yields, after some manipulations, to 
\begin{equation*}
\frac{\partial f^\veps(\tau,x)}{\partial \tau}+\frac{ j^\veps(\tau,x-\veps)f^\veps(\tau,x-\veps) - j^\veps(t,x)f^\veps(t,x)}{\veps}=o(1)\,,
\end{equation*}
with $j^\veps(\tau,x)=x^\alpha\left(u^\veps(\tau)-\frac{q}{x^\gamma}\right)$. Formal arguments lead, as $\veps \to 0$, to a solution $f$ of Eq. \eqref{eq:LS}. In turns, the mass conservation \eqref{eq:BD_mass} becomes
\begin{equation*}
\rho =\sum_{i=1}^\infty i c_i(\tau) = \sum_{i=1}^{i_\veps} i c_i(\tau)+\sum_{i=i_\veps}^\infty ic_i = \rho_s + \int_0^\infty x f(\tau,x) \, dx + o(1)\,,
\end{equation*}
At the limit, we obtain \smash{$\int_0^\infty x f(\tau,x) \, dx =\rho-\rho_s$} which measures large clusters formation. Such condition, complemented with the LS equation \eqref{eq:LS}, allows to determined $u$ by the following expression
\begin{equation*}
u(\tau)=\frac{q\int_0^\infty x^{\alpha-\gamma}f(\tau,x)\,,dx}{\int_0^\infty x^{\alpha}f(\tau,x)\,dx}\,.
\end{equation*}
We now state the result obtained by  \cite{Niethammer2003}.
\begin{theorem}[Lifschitz-Slyozov-Wagner limit, \cite{Niethammer2003}]
Assume kinetic coefficient are given by Eq.~\eqref{eq:example_coefficients}, that the initial condition $c^\veps(0)$ satisfies $H(c^\veps(0)|c^{\rho_s})=\veps^\gamma$ and that $\sum_{i\geq M/\veps} i c_i^\veps(0) \to 0$ as $M$ goes to infinity uniformly in $\veps>0$. 

There is a subsequence $\{\veps_n\}$ converging to $0$,  a measure-valued function $t\mapsto \nu_t$ solution of LS equation \eqref{eq:LS} in $\mathcal D'(\Rb_+\times(0,+\infty))$ such that
\begin{equation*}
\int_0^\infty \vphi(x) f^\veps(\tau,x)\, dx \to \int_0^\infty \vphi(x) \nu_t(dx)\,,
\end{equation*}
locally uniformly in  $t\in\Rb^+$, for all $\vphi \in C_0^0((0,\infty))$ and for all $t\geq 0$, and
\[ \int_0^\infty x \nu_t(dx) = \rho-\rho_s\,.\]
\end{theorem}
We also mention the case of vanishing small excess of density, $\rho-\rho_s\to 0$ as $\veps\to 0$, by \cite{Niethammer2004}, where the authors recovered the LS equation, in a similar framework.
\subsubsection{Rescaled solution of BD for large monomer density}

Another point of view is to consider fast reaction rates $a_ic_1 c_i\sim b_{i+1}c_{i+1}$ of order $1/\veps$, where $0<\veps \ll 1$, together with a large excess of monomers. Namely, the characteristic number of free particles $c_1$ is two orders of magnitude greater than the characteristic number  of clusters with size $i\geq 2$. Following \cite{Collet2002}, alternatively \cite{Deschamps2016}, this leads to a rescaled version of the BD equations~(\ref{sys:BD-1}-\ref{sys:BD-2}) given, for $\veps>0$, by
\begin{align}
 & \frac{d}{dt} u^\veps  =   -  \veps J_{1}^\veps -  \veps\sum_{i \geq {1}} J_i^\veps \,, \label{sys:epsBD-1}\\
 & \frac{d}{dt} c_{i}^\veps  =  \frac{1}{\veps}\Big{[} J_{i-1}^\veps -  J_i^\veps \Big{]}\,,  \label{sys:epsBD-2}
\end{align}
for every $i\geq 1$,  where $u^\veps$ is the dimensionless version of $c_1$ (not to be confused with the previous section) and the scaled fluxes are
\begin{equation*}
 J_{1}^\veps =  \alpha^\veps (u^\veps)^2- b_2^\veps c_{2}^\veps\,,\qquad J_{i}^\veps = 
a_i^\veps u^\veps c_i^\veps - b_{i+1}^\veps c_{i+1}^\veps\,, 
\end{equation*}
for every $i\geq 1$. Theorem \ref{thm:BD_wellposed} provides existence and uniqueness of solution at fixed $\veps>0$. \cite{Collet2002} constructed a sequence of ``density'' approximations in the Lebesgue space $L^1(\Rb_+)$ by, for all $t\geq0$ and $x\geq 0$
\begin{equation*}
 f^\veps(t,x) = \sum_{i\geq 2} c_i^\veps(t)\indic{\Lambda_i^\veps}(x)\,, 
\end{equation*}
where $\Lambda_i^\veps=[(i-1/2)\veps, (i+1/2)\veps)$ for each $i\geq 2$. Note the first cluster is excluded from the density, it is like assuming a solute with density $f^\veps$ belonging to the solvent $u^\veps$ (in large excess). Then, macroscopic aggregation and fragmentation rates are constructed as functions on $\Rb_+$ (similarly to $f^\veps$), for each $\veps>0$ and  $x\geq 0$,
\begin{equation*} 
  a^\veps(x)  = \ds  \sum_{i \geq 2} a_i^\veps \indic{\Lambda_i^\veps}(x)\,, \qquad b^\veps(x) =   \sum_{i \geq 2} b_i^\veps \indic{\Lambda_i^\veps}(x)\,.
\end{equation*}
This scaling supposes the first coagulation rate $\alpha^\veps$ is faster (order $1/\veps^2$) than the other rates $a_i^\veps$ for $i\geq 2$, which justifies the use of another notation $\alpha^\veps$ and a special treatment outside the function $a^\veps$. Theoretical justifications can  be found in \cite{Collet2002}. Finally, the balance of mass reads in this case, for all $t\geq 0$
\begin{equation} \label{eq:epsBD_mass}
 u^\veps(t) + \int_0^\infty x f^\veps(t,x)\, dx = \rho^\veps\,,
\end{equation}
for some $\rho^\veps>0$. The value of $\rho^\veps$ is entirely determined by the initial condition at time $t=0$.

Again we deal with the limit $\veps\to 0$, and we hope the limit of $f^\veps$ satisfies in some sense the LS equation \eqref{eq:LS}. Let us introduce few hypotheses for  the limit theorem, namely we assume, there exists a constant $K>0$, independent on $\veps>0$, such that, for all $x\geq 0$, 
\begin{equation}\label{eq:bound_coeff}
a^\veps(x) + b^\veps(x) \leq K(1+x)\,.
\end{equation}
Also, we assume there exists a measure $\mu^{in}$ on $\Rb_+$ such that
\begin{equation} \label{eq:cv_init}
 \lim_{\veps\to 0} \int_0^\infty \vphi(x) f^\veps(0,x) \, dx = \int_0^\infty \vphi(x) \mu^{\rm in}(dx)\,,
\end{equation}
for all $\vphi\in \Cc_0((0,+\infty))$ and 
\begin{equation} \label{eq:uniform_int}
 \lim_{R\to+\infty} \, \sup_{\veps>0} \, \int_R^\infty x f^{\veps}(0,x)\, dx = 0\,.
\end{equation}
This estimate on the  tail of the initial distribution is a classical argument which increase the compactness and will allow then to pass to the limit in the balance of mass \eqref{eq:epsBD_mass}.
%
%

Finally, we resume in the following the results obtained by \cite{Collet2002} in their Theorem 2.3, by \cite{Laurencot2002a} in Theorem 2.2 for a different framework, and also modified by \cite{Deschamps2016}, in Lemma 5.
\begin{theorem}[Lifschitz-Slyozov limit, \cite{Collet2002}, \cite{Laurencot2002a}]\label{thm:BDtoLS}
Assume that $\alpha^\veps$ is uniformly bounded, and that $a^\veps$ and  $b^\veps$ satisfy Eq.~\eqref{eq:bound_coeff}. Suppose moreover that there exists $\rho\geq 0$  and two non-negative real functions $a$ and $b$ defined on $\Rb_+^*$ such that, when $\veps$ converges to $0$, $\rho^\veps$ converges to $\rho$, $a^\veps$ and $b^\veps$ converge locally uniformly on $\Rb_+^*$ toward, respectively, $a$ and $b$. 

If the family $\{f^{\veps}(0,\cdot)\}$ satisfies Eqs. \eqref{eq:cv_init} and \eqref{eq:uniform_int}, then from all sequences $\{\veps_n\}$ converging to $0$ we can extract a subsequence still denoted  $\{\veps_n\}$ such that 
\begin{equation} \label{eq:cv_feps}
 \lim_{n\to \infty} \int_0^\infty \vphi(x) f^{\veps_n}(t,x) \, dx = \int_0^\infty \vphi(x) \mu(t,dx)\,,
\end{equation}
locally uniformly in $t\in \Rb_+$, and for all $\vphi\in\Cc_0((0,+\infty))$, where $\mu := (\mu(t,\cdot))_{t\geq0}$ is a measure-valued function satisfying the LS equation \eqref{eq:LS} in $\mathcal D'(\Rb_+\times(0,+\infty))$ where $u\in\Cc(\Rb_+)$ is non-negative and satisfies, for all $t\geq 0$,
\begin{equation} \label{eq:masscons_final}
  u(t) + \int_0^\infty x \mu(t,dx) = \rho\,.
\end{equation}
\end{theorem}
The proof relies, mainly, on moment estimates and equicontinuity arguments. This theorem does not conclude on the full convergence of the family as $\veps \to 0$. To that it requires a uniqueness argument of the limit problem Eq.~\eqref{eq:LS} in measure with the balance of mass \eqref{eq:masscons_final}. Looking Eq.~\eqref{eq:LS} against functions in $\mathcal D(\Rb_+\times (0,+\infty))$ allows uniqueness with the necessary condition that the flux $j(t,x)$ points outward the domain at $x=0$ (for instance if $a(0)\rho-b(0)<0$). We refer to the works by  \cite{Niethammer2000}, \cite{Collet2000} and by \cite{Laurencot2001} for the well-posedness theory on the Lifshitz-Slyozov equation. Also, we mention that the convergence in Eq. \eqref{eq:cv_feps} has also been shown to hold in a functional density space, in $L^1(xdx)$, by \cite{Laurencot2002}.

We are now concerned with the case the flux $j(t,x)$ points inward the domain at $x=0$, for instance if $a(0)u(0)-b(0)>0$, or more generally if the characteristics, backward solution of  
\[\frac{d}{dt} x = j(t,x)\]
goes back to $x=0$ in finite time. In this case, it is hopeless to obtain a well-defined limit to the LS equation \eqref{eq:LS} without a boundary condition, of type \eqref{eq:BC_LS}. A rigorous identification of the boundary condition has been performed by \cite{Deschamps2016}. It was obtained  through the limit of the rescaled BD equations (\ref{sys:epsBD-1}-\ref{sys:epsBD-2}) in the spirit of Theorem \ref{thm:BDtoLS}. More precisely, we assumed, $a(x)\sim_0 \bar a x^{r_a}$ and $b(x)\sim_0 \bar b x^{r_b}$ with $r_a \leq r_b$ and $r_a<1$. These assumptions allow a fine control of the pointwise value of the solution at $x=0$ to obtain the boundary value. The limit obtained is a measure-valued solution to LS on $[0,T]$, identifiable if $\sup_{t\in[0,T]} u(t) > \lim_{x\to0} b(x)/a(x)$ which correspond to time interval on which characteristic goes back to $x=0$. Let us present an informal version of a result we obtained. 
\begin{theorem}[Boundary value, \cite{Deschamps2016}]
 A ``good'' boundary condition at $x=0$ for the Lifschitz-Slyozov equation, when $a(x)= \bar a x^{r_a}$ and $b(x)=\bar b x^{r_b}$ with $r_a<1$ and $r_a\leq r_b$, is
 
 \[\lim_{x\to 0^+} j(t,x) f(t,x) = \begin{cases}
                                                 \alpha u(t)^2\,, & \textup{if}\ r_a < r_b\,, \ u(t)>0 \,;\\
                                                 \frac{\alpha}{\overline a} u \left(\overline{a}u-\overline{b}\right)\,, & \textup{if}\ r_a=r_b\,, \ u(t) > \bar b / \bar a\,,
                                                \end{cases}
\]
where $\alpha$ is the limit of $\alpha^\veps$ as $\veps$ goes to $0$. In both cases, this also reads 
\[ \lim_{x\to 0^+} x^{r_a} f(t,x) = \frac{\alpha}{\overline a} u(t)\,. \]
\end{theorem}
Note the conditions on $r_a$, $r_b$ and $u$ are well related to incoming characteristic. Theorems 1 and 2 by \cite{Deschamps2016} also assumed a technical growth condition (in $\veps$) on the ``relatively small'' sizes, through the condition
\begin{equation*}
  \sup_{\veps>0} \, \sum_{i\geq 2} \veps^{r_a} c_i^{\rm in,\,\veps} e^{-iz} < +\infty\,, 
\end{equation*}
for all $z\in(0,1)$. This is the key estimates which is proved to propagate in time (see Proposition 2). This allows a quasi steady-state limit of the small cluster concentrations, that behave as  fast variables in Eq.~\eqref{sys:epsBD-2}. Note in the case of exact power law, \cite{Deschamps2016} also proved with extra reasonable assumptions on initial conditions, that the limit measure solution has a density with respect to $x^{r_a} dx$. Finally, other scalings of the first fragmentation rate are investigated by \cite{Deschamps2016}. Also, these results do not provide a complete answer. Indeed, uniqueness for the inward case is not achieved and we are not aware if $u$ can cross the threshold $\lim_{x\to0} b(x)/a(x)$.
%
%
%

\begin{remark} \label{rmk:second_order}
Second-order approximations (Fokker-Planck like) of BD equations are still under intense active research, and a full satisfactory answer is still an open problem, see proposed equations by \cite{Velazquez1998,Velazquez2000}, \cite{Hariz1999,Collet2002,Collet2004}, \cite{Conlon2016}. Arbitrary higher order terms are formally derived by \cite{Niethammer2003}.
\end{remark}

\subsection{Time-dependent properties, Metastability and Classical nucleation theory}\label{sec:metastability}

The following properties are of the most important ones in application of the BD equations to phase transition. Yet, as for the convergence rate to equilibrium, coarsening and evolution of large sized clusters, available results are still incomplete. 
The main result we are aware of on metastability for BD equations (\ref{sys:BD-1}-\ref{sys:BD-2}) is given by \cite{Penrose1989}. The ideas of classical nucleation theory goes back to \cite{Becker1935}, and is built on the remark that there exist steady-state solutions of the Eqs.~\eqref{sys:BD-2} (with $c_1$ constant) with non-zero steady-state flux, which can be arbitrary small in some sense. This very small steady-state flux is interpreted as the rate of formation of larger and larger cluster, leading to a phase transition phenomena in long time. The term metastability in such theory refers to the fact that the rate is arbitrary small. \cite{Penrose1989} goes much beyond by extending this notion of metastability to a time-dependent phenomenon (instead of a steady-state one). Indeed, he could exhibit a solution of the full system (\ref{sys:BD-1}-\ref{sys:BD-2}) that enters a state that lived for exponentially long time, yet can be distinguished from the equilibrium state. This solution is a super-critical solution, with $\rho>\rho_s$, and is required to have a well prepared initial condition. This solution is also related in some sense to an extremely small common flux value. It remains an important open question to know whether the metastable state can be reached from a larger class of initial data. 

\cite{Penrose1989} considered technical conditions on coefficients which are essentially satisfied by the ones given by Eq.~\eqref{eq:example_coefficients}. The crucial initial condition is then constructed as follows. For any $z>z_s$, let $f_i(z)$ be the unique solution of 
\begin{equation*}
a_{i-1}zf_{i-1}(z)-(b_i+a_iz)f_i(z)+b_{i+1}f_{i+1}(z)=0\,,\quad i\geq 2\,,
\end{equation*} 
with end conditions $f_1(z)=z$ and $\sup_i f_i(z)<\infty$. Actually, $f_i$ can be solved explicitly by (for $z>z_s$ the reader can check that the infinite series are convergent)
\begin{equation*}
f_i(z)=J(z)Q_iz^i\sum_{r=i}^\infty\frac{1}{a_iQ_iz^{i+1}}\,,\quad J(z):=\left[\sum_{r=1}^\infty\frac{1}{a_iQ_iz^{i+1}}\right]^{-1}\,.
\end{equation*}
Let $i^*$ be the critical cluster size defined as the (unique) size that minimizes the quantity $a_iQ_iz^i$. The metastable state exhibited by \cite{Penrose1989} has to be understood in the limit of small excess of density, $z\searrow z_s$. The following terminology is used
\begin{itemize}
	\item $g(z)$ is exponentially small	 if for each $m>0$, $g(z)=O\left((z-z_s)^m\right)$.
	\item $g(z)$ is at most algebraically large if for some $m>0$, $g(z)=O\left((z-z_s)^{-m}\right)$.
\end{itemize}
The main theorem by \cite{Penrose1989} reads
\begin{theorem}[Metastability, \cite{Penrose1989}]
Let $c$ be the solution of the BD Eqs.~\eqref{sys:BD-1}-\eqref{sys:BD-2} with initial condition
\begin{equation*}
c_i(0)=\begin{cases}
f_i(z)\,,\quad \text{ if } \quad i\leq i^*\,,\\
Q_iz_s^i\,,\quad \text{ if } \quad i> i^*\,.
\end{cases}
\end{equation*}
Then $c$ has an exponentially long lifetime as $z\searrow z_s$, in the sense that for each fixed $i$ (note that $i^*\to\infty$):
\begin{itemize}
	\item if $t$ is at most algebraically large, then $c_i(t)-c_i(0)$ is exponentially small
	\item $\lim_{t\to\infty}[c_i(t)-c_i(0)]$ is not exponentially small
\end{itemize}
\end{theorem}
\noindent Thus, cluster with size $i\ll i^*$ remain exponentially close to their initial values, until an exponentially long time has elapsed. But eventually they do change. Note that the initial values for the small clusters, $f_i(z)$, correspond to the steady-state values of the classical nucleation theory, for which $J_{i-1}(0)=J_i(0)$ for all $2\leq i<i^*$, and the common flux value is $J(z)$, which is also exponentially small as  $z\searrow z_s$. We refer the reader to \cite[Theorems 1 and 2]{Penrose1989} for orders of magnitude of $i^*$, $J(z)$ and quantification of the (small) growth rate of large clusters of size greater than $i^*$.

\begin{remark}
	The numerical illlustration of the metastability is a problem \textit{per se}, we refer the reader to the two nice papers by \cite{Carr1995} and by \cite{Duncan2001}, where numerical schemes are derived and are shown to consistently represent the metastable states. The reader may also look at the section \ref{ssec:meta_stoc} where numerical simulations of the stochastic Becker-D\"oring are shown. Finally, let us mention that analogous metastability properties have been investigated in the classical linear version of  BD by \cite{Penrose1989} and \cite{Kreer1993}, in a finite-dimensional truncated version by \cite{Dunwell1997} and \cite{Duncan2002}, and in a thermodynamically consistent version of the BD system by \cite{Ssemaganda2013}.
\end{remark}
%


%
%
%
%
%

\section{Stochastic Becker-D\"oring model}

Due to space considerations, we will not detail historical facts on the study of stochastic coagulation-fragmentation models. Let us just mention that the first study of a stochastic coagulation models is widely attributed to \cite{Marcus1968} and \cite{Lushnikov1978} which give the name to the Marcus-Lushnikov process, stochastic analog of the pure coagulation Smoluchowski’s equations. Up to our knowledge, \cite{Whittle1965} and \cite{Kelly1979} are pioneering in the study of more general stochastic coagulation-fragmentation models (including the  Becker-D\"oring model). See \cite{Aldous1999} and the discussion in \cite{Freiman2005} for more details.

\subsection{Definition and State-space}

A stochastic version of the Becker-D\"oring model may be defined as a continuous time Markov chain analog of the set of ordinary differential equations (\ref{sys:BD-1}-\ref{sys:BD-2}), for which transition are given by the same set of kinetic reactions \eqref{eq:kinetic_reaction}, but modeling \textit{discrete numbers} of clusters instead of continuous concentrations. Precisely, given a positive integer $M$, we define the state space
\begin{equation*}
X_M:= \left\lbrace C=(C_i)_{i\geq 1} \in \Nb^\Nb\, :\, \sum_{i=1}^{M}iC_i = M\right\rbrace\,.
\end{equation*} 
On $X_M$, we introduced the following operators defined by, for any configuration $C$ on $X_M$,
\begin{equation*}
\begin{array}{rcl}
\ds R^+_1 C &  \ds =  & \ds(C_1-2\,,C_2+1\,,\cdots\,,C_i\,,\cdots) \\
\ds R^-_2 C &  \ds =  & \ds(C_1+2\,,C_2-1\,,\cdots\,,C_i,,\cdots)
\end{array}
\end{equation*} 
and, for any $i\geq 2$,
\begin{equation*}
\begin{array}{rcl}
\ds R^+_i C &  \ds =  & \ds(C_1-1\,,C_2\,,\cdots\,,C_i -1\,,C_{i+1}+1\,,\cdots) \\
\ds R_{i+1}^- C &  \ds =  & \ds(C_1+1\,,C_2\,,\cdots\,,C_i +1\,,C_{i+1}-1\,,\cdots)
\end{array}
\end{equation*} 
Given non-negative kinetic rates $(a_i)_{i\geq 1}$, $(b_i)_{i\geq 2}$, the stochastic Becker-D\"oring model (SBD) is defined as the continous time Markov chain on $X_M$ with transition rates
\begin{equation*}
\left\lbrace\begin{array}{rclr}
\ds q(C,R^+_1 C) &  \ds =  & a_1C_1(C_1-1)\,, &\\
\ds q(C,R^+_i C) &  \ds =  & a_iC_1C_i\,,& i\geq 2\,, \\
\ds q(C,R^-_i C) &  \ds =  & b_iC_i \,,& i\geq 2\,.
\end{array}\right.
\end{equation*} 
Given an initial configuration $C^{\rm in}\in X_M$ (deterministic or random), the configuration $C(t)$ defined by the SBD may alternatively be represented as the solution of the following system of stochastic equations  
\begin{equation}\label{HOMOEQN0}
\left\lbrace 
\begin{array}{rcl}
\ds C_{1}(t) & \ds = & \ds C_{1}^{\rm in} -2J_1(t)-\sum_{i\geq 2} J_i(t)\,,  \\
\ds C_{i}(t) & \ds = & \ds C_{i}^{\rm in} + J_{i-1}(t)-J_i(t)\,,\quad i\geq 2\,, \\
\end{array}
\right.
\end{equation}
with 
\begin{equation*}
J_{i}(t)= Y_i^+\Big{(}\int_0^t a_i C_{1}(s)(C_{i}(s)-\delta_{1,i} )ds\Big{)}  \ds -Y_{i+1}^-\Big{(}\int_0^t b_{i+1} C_{i+1}(s)ds\Big{)}\,,\quad i\geq 1\,,
\end{equation*}
where $\delta_{1,i}=1$ if $i=1$ and $\delta_{1,i}=0$ if $i>1$ and $Y_i^+$, $Y_{i+1}^-$ for $i\geq 1$ are independent standard Poisson processes. Analogy between Eq.~\eqref{HOMOEQN0} and Eq.~(\ref{sys:BD-1}-\ref{sys:BD-2}) is clear. The number of clusters of size $i\geq 2$ evolves according to the differences between two (stochastic) cumulative counts $J_{i-1}$ and $J_{i}$. Finally, we may also identified the SBD with the help of its infinitesimal generator $L_M$, defined by, for any bounded functions $f$ on $X_M$,
\begin{equation*}
L_Mf(C)=\sum_{i=1}^{M-1} \left[f(R^+_i C)-f(C)\right]a_i C_{1}(C_{i}-\delta_{1,i} ) + \left[f(R^-_{i+1} C)-f(C)\right]b_{i+1}C_{i+1}\,.
\end{equation*}
Thanks to the Markov processes theory, we deduce in particular that, for any bounded functions $f$ on $X_M$, 
\begin{equation*}
f(C(t))-f(C^{\rm in})-\int_0^t L_Mf(C(s))ds
\end{equation*}
is a centered martingale, and, taking $f_C(C')=\indic{\{C'=C\}}$, we deduce the following Backward kolmogorov equation on the probability $P(t,\cdot)$ on $X_M$ (Master equation)
\begin{multline}\label{eq:mastereq}
\frac{d}{dt}P(t;C)= \sum_{i=1}^{M-1}  a_{i}(C_1+1)(C_{i}+1+\delta_{1,i})P(t;R_{i+1}^- C)- a_i C_{1}(C_{i}-\delta_{1,i} )P(t;C) \\
 + \sum_{i=2}^{M}  b_{i}(C_{i}+1)P(t;R^+_{i-1} C)- b_i C_{i}P(t;C)\,.
\end{multline} 

Although the well-posedness of the SBD model is of course standard (as a pure-jump Markov process on a finite state-space), a first non trivial question arises with respect to the precise description of the state space, and in particular to its cardinality. In fact, the state space $X_M$ is given by all possible partitions of the integer $M$, a well-known problem in combinatorics. In particular, one can show the recurrence formula and the asymptotic as $M\to\infty$, \cite[chap I.3]{Flajolet2009}\footnote{R.Y thanks Bence Melykuti for pointing out this fact}
\begin{equation*}
M \mid X_M \mid = \sum_{i=1}^M \sigma(i)\mid X_{M-i}\mid\,,\quad \mid X_M \mid \propto \frac{1}{4M\sqrt{3}}\exp\left(\pi \sqrt{\frac{2M}{3}}\right)\,,
\end{equation*}
where $\sigma(i)$ is the sum of the divisors of $i$ (e.g. $\sigma(6)=1+2+3+6=12$).

\begin{remark}
We mention that some terminology in the literature may be confusing. Indeed, some authors \cite{Bhakta1995} have named the deterministic Becker-D\"oring system (\ref{sys:BD-1}-\ref{sys:BD-2}) a stochastic version of the Lifshitz-Slyozov(-Wagner) equations. Such terminology seems to be motivated by the fact that the size of clusters are modeled as discrete variable in Eq.~(\ref{sys:BD-1}-\ref{sys:BD-2}), and that such system has the "flavor" of a master equation for a random walk in $\Nb^+$. 
\end{remark}

\begin{remark}
As for the BD system~(\ref{sys:BD-1}-\ref{sys:BD-2}), some variants have been considered for the SBD. Let us mention for instance the constant monomer system studied in \cite{Yvinec2016} (which leads to a Poissonian equilibrium distribution), the exchange-driven growth model in \cite{Ben-Naim2003} (where clusters exchange monomer in one single step), some reduced version for specific kinetic rates ($a(i)=i, b(i)=0$) see \cite{Doumic2016} and \cite{Eugene2016} or for fixed number of clusters (only one or two clusters can be present) \cite{Penrose2008}, \cite{Rotstein2015} and \cite{Yvinec2016}. Of  course, the SBD can be seen as a particular case of more general coagulation-fragmentation processes  \cite{Bertoin2006}. However, due to its specificity, it seems that some of the results available on general coagulation-fragmentation processes are not straightforwardly applicable (or do not bring interesting conclusions). Finally, although out of the scope of this survey, let us mention the interesting links between the (stochastic) BD system with some lattice models \cite{Chau2015}, \cite{Dehghanpour1997}, \cite{Hollander2000}, \cite{Bovier2000}, \cite{Ercolani2014}, in particular for nucleation and phase transition.
\end{remark}

\subsection{Long-time behavior}

Although the equation \eqref{eq:mastereq} is linear with respect to $P(t,\cdot)$, the size of the state space being exponentially large as $M\to \infty$, it is illusory to obtain a full exact solution of Eq.~\eqref{eq:mastereq}. Yet, perhaps surprisingly, the stationary solution of Eq.~\eqref{eq:mastereq} has a relatively simple form, namely a product-form \cite{Anderson2010}. Indeed, the (unique) stationary probability $\Pi$ on $X_M$ of Eq.~\eqref{eq:mastereq} is given by \cite[Theorem 8.1]{Kelly1979}
\begin{equation}\label{eq:stat_sbd}
\Pi(C)=B_M \prod_{i=1}^{M}\frac{(Q_i)^{C_i}}{C_i !}\,,
\end{equation}
where  $B_M$ is a normalizing constant and $Q_i$ is defined by Eq.~\eqref{eq:equilibrium}. One may verify simply that the following detailed balance condition holds \cite[Theorem 1.2]{Kelly1979} 
\begin{equation*}
\Pi(C)q(C,R_i^+C)=\Pi(R_i^+C)q(R_i^+C,C)
\end{equation*}
Note also that, for all $z>0$, with $B_z:=B_M/z^M$, the expression \eqref{eq:stat_sbd} may be rewritten $\Pi(C) = B_z \prod_{i=1}^{M}\frac{(Q_iz^i)^{C_i}}{C_i !}$, which has a clearer analogy with the deterministic equilibrium of the BD equation. Finally, the distribution $\Pi$ has the following probabilistic meaning: let $Z_i$, $i=1,\cdots,M$, be independent Poisson random variables with respective means $Q_i$, then it is easily seen that, for all $C\in X_M$,
\begin{equation*}
\Pi(C)=\mathbf P \left\{Z_1=C_1,\cdots,Z_M=C_M \mid \sum_{i=1}^M iZ_i = M\right\}\,.
\end{equation*}
For the stationary distribution $\Pi$, the expected number of clusters of size $i$ is 
\begin{equation*}
\mathbf E_\Pi  C_i=Q_iB_M/B_{M-i}\,,
\end{equation*} 
and the probability that a randomly chosen particle lies in a cluster of size $i$ is  $iQ_iB_M/MB_{M-i}$, from which we deduce that the normalizing constant $B_M$ satisfies the recursive formula (with $B_0=1$)
\begin{equation}\label{eq:cst_recurisv}
MB_M^{-1}=\sum_{i=1}^M iQ_iB_{M-i}^{-1}\,.
\end{equation}
Moreover, $B_M^{-1}$ is the coefficient of $z^M$ in the power series expansion of
\begin{equation*}
G(z)=\exp(\sum_i Q_i z^i)
\end{equation*}
\begin{remark}
In some examples, the recursive formula \eqref{eq:cst_recurisv} may be solved exactly. For instance, if $a_i = a i$, $b_i = b i$, then the equilibrium probability is given by the closed-form formula
\begin{equation*}
\Pi(C)= \left(\begin{array}{c}b/a + M-1 \\ M\end{array}\right)^{-1} \prod_{i=1}^{M}\frac{1}{C_i !}\left(\frac{b}{ai}\right)^{C_i}\,,
\end{equation*}
\end{remark}
Besides the analytical form of the equilibrium distribution $\Pi$, it is a natural question to ask what is its limiting behavior as $M\to\infty$. Under the assumption that
\begin{equation}\label{eq:hyp_eq_sto}
\lim_{i\to\infty} \frac{a_i}{b_{i+1}}=z_s>0\,,
\end{equation}
(which is slightly stronger than the hypothesis on $Q_i^{1/i}$ used in Theorem \ref{thm:H-theorem} and \ref{thm:convergence_equilibrium}), one can show \cite{Freiman2002,Bell2003} that $G$ has also for radius of convergence $z_s$, and
in such case, the expected number of clusters of size $i$ has a limit as $M\to\infty$, given by 
\begin{equation}\label{eq:lim_mean}
\lim_{M\to\infty} \mathbf E_\Pi C_i =Q_iz_s^i\,,
\end{equation} 
Other functionals of the stationary distribution $\Pi$ have been derived by \cite{Durrett1999}. In particular, let us mention that the variance of $C_i$, under $\Pi$ and with hypothesis \eqref{eq:hyp_eq_sto}, satisfies the same asymptotic relation \eqref{eq:lim_mean}, and that $C_i,C_j$, $i\neq j$, becomes asymptotically uncorrelated as $M\to\infty$. It is also interesting to note the link of the limit \eqref{eq:lim_mean} with the supersaturation case in the deterministic BD theory, see Theorem \ref{thm:convergence_equilibrium}. Study of the limit shape of the stationary distribution $\Pi$ (and quantities like the size of the largest or lowest component) is a well-known problem in statistical physics or in combinatorics (study of random integer partitions and Young diagrams) and goes back to Khinchin's probabilistic method \cite{Khinchin1960}. Detailed description of such field is out of the scope of this survey, and we refer the reader to \cite{Freiman2005}, \cite{Erlihson2008}, \cite{Han2008}, \cite{Granovsky2013}, \cite{Ercolani2014} for recent results.

In contrast to the deterministic theory, we are not aware of any work quantifying the speed of convergence toward the equilibrium distribution~\eqref{eq:stat_sbd} (which has to be exponential). In particular, it would be interesting to study how this rate behave as $M\to \infty$.
 

\begin{remark}
Strong binding limit for constant coefficients has been considered in \cite{DOrsogna2012} (linked to the almost pure-coagulation deterministic dynamics in \cite{King2002}) and illustrates how mass incommensurability arises for finite mass $M$, when a fixed maximal cluster size $N<M$ is further imposed.
\end{remark}

\subsection{Large Number and relation to deterministic Becker-D\"oring }

A first natural question when comparing the SBD and the BD system, is that can we recover the deterministic equations in the limit $M\to+\infty$? The main tool to answer such question is the tightness of stochastic processes, which provides an appropriate compactness property for a sequence of rescaled solutions of the SBD. As a particular case, \cite{Jeon1998} has considered the sequence of stochastic processes $\{C^n(t)\}$ in \smash{$X^{+}_n:=\left\{\frac{1}{n}C\,:\, C \in \Nb^\Nb \,,  \, \sum_{i\geq 1} i C_i =n \right\}\subset X^{+}\subset l^2 $}, defined by the generator
\begin{equation}\label{eq:sbd_rescaled}
L^nf(C)=n\sum_{i=1}^{n} \left[f(R^+_{i,n} C)-f(C)\right]a_i C_{1}(C_{i}-\delta_{1,i} ) + \left[f(R^-_{i+1,n} C)-f(C)\right]b_{i+1}C_{i+1}\,,
\end{equation}
where, for all $i\geq 1$,
\begin{equation*}
\begin{array}{rcl}
\ds R^+_{i,n} C &  \ds =  & \ds(C_1-1/n\,,C_2\,,\cdots\,,C_i -1/n\,,C_{i+1}+1/n\,,\cdots) \\
\ds R_{i+1,n}^- C &  \ds =  & \ds(C_1+1/n\,,C_2\,,\cdots\,,C_i +1/n\,,C_{i+1}-1/n\,,\cdots)
\end{array}
\end{equation*} 
Under such classical scaling (which satisfies the system size expansion), one can prove
\begin{theorem}[Law of Large Numbers, \cite{Jeon1998}]
If $a(i),b(i)$ are such that
\begin{equation}\label{eq:hyp_tension}
\sup_{C \in X^+: \sum{iC_i\leq 1}} \,\sum_{i\geq 1} a(i)C_i<\infty\,,\quad \sup_{C \in X^+: \sum{iC_i\leq 1}} \,\sum_{i\geq 1} b(i)C_i<\infty\,,
\end{equation}
then the laws of the stochastic process $\{C^n(t)\}$ defined by Eq.\eqref{eq:sbd_rescaled} form a tight sequence as a \textit{c\`adl\`ag} process in $l^2$.
\end{theorem}
Note that hypothesis \eqref{eq:hyp_tension} is trivially satisfied for sublinear function of $i$. Also, it is clear that any weak limit of $\{C^n(t)\}$ is a solution of the BD system (\ref{sys:BD-1}-\ref{sys:BD-2}), which is an alternative proof of existence of solution of the BD system. Finally, convergence of the whole sequence may be obtained with the uniqueness result stated in Theorem \ref{thm:BD_wellposed}.

\noindent We are not aware of any rigorous derivation of a second-order approximation of such limit, which should reasonably be a langevin stochastic differential equation version of the BD system (\ref{sys:BD-1}-\ref{sys:BD-2}).

\subsection{Time-dependent properties, Metastability and stochastic nucleation theory}\label{ssec:meta_stoc}

\begin{figure}[!h]
\begin{center}
\includegraphics{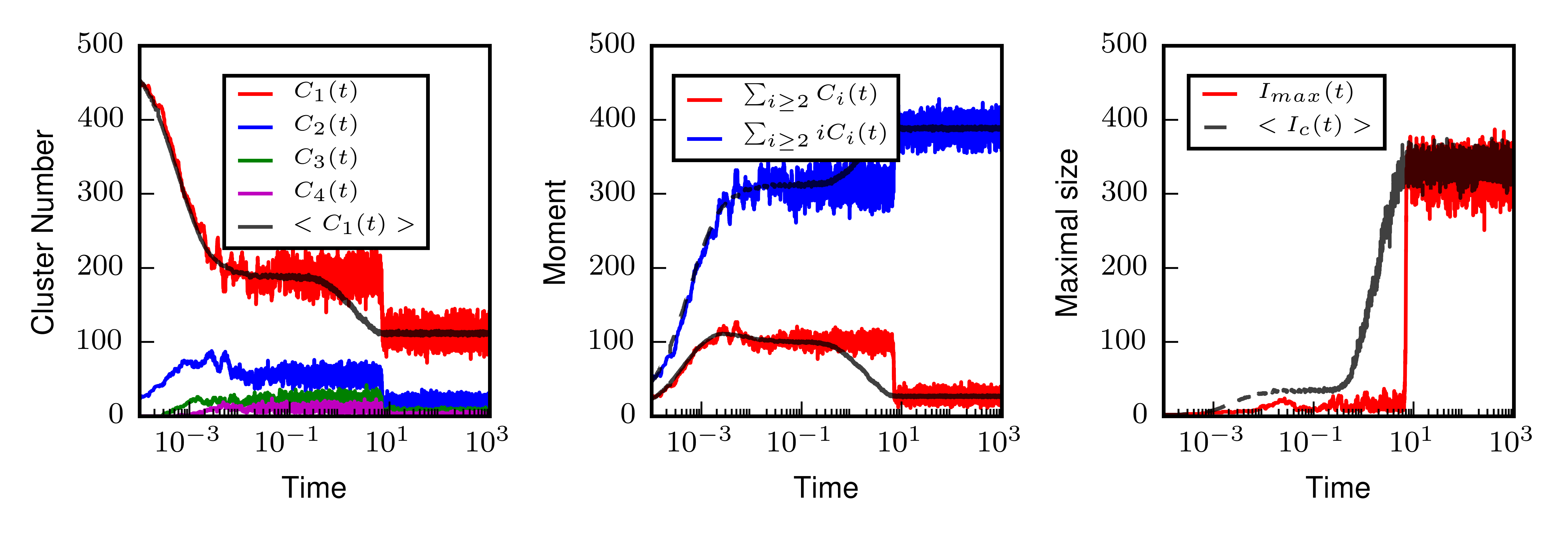}
\end{center}
\caption{Time trajectories of the SBD system \eqref{HOMOEQN0} with kinetic coefficient given by \eqref{eq:example_coefficients}, with $M=500$, $\alpha=2/3$, $\gamma=1/3$, $z_s=500/11$, $q=10/11$. On the left, we plot a stochastic realization of the number of Monomers, Dimers, Tri-mers and $4$-mers, together with the sampled average over $100$ realizations for the number of Monomers. On the middle, we plot the total mass in clusters and their numbers, and on the right, the maximal cluster size}\label{fig1}
\end{figure}

\begin{figure}[!h]
\begin{center}
\includegraphics{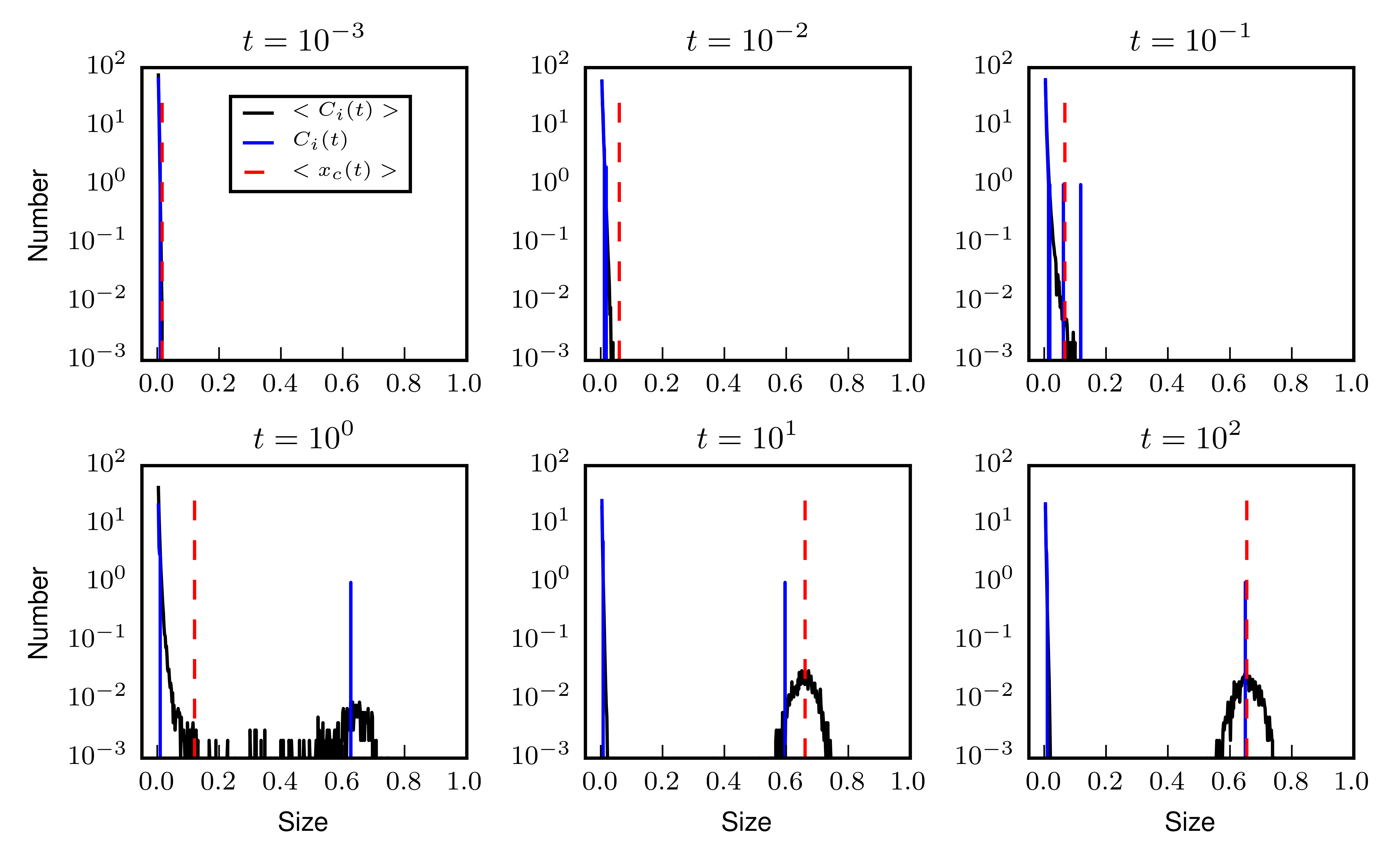}
\end{center}
\caption{Cluster size distribution at distinct times $t$, corresponding to Fig \ref{fig1}. The sizes are rescaled by $M$, that is  $x=2/M,\cdots,1$. In blue we plot the distribution of a stochastic realization, in black we represent the sampled averaged distribution over $1000$ realizations, and in red we plot the (rescaled) critical size $x_c=I_c/M$.}\label{fig2}
\end{figure}

Up to our knowledge, the early work \cite{Schweitzler} paves the way to study fluctuations of the time-dependent cluster
distributions and first passage time in stochastic finite system nucleation models. Using physical arguments, they investigated reaction rates of the form $a_i \approx i^{2/3}$ and $b_i \approx i^{2/3}y_0e^{qi^{-1/3}}$, which are asymptotically similar to Eq.~\eqref{eq:example_coefficients} (which $\alpha=2/3$, $\gamma=1/3$). One can notice that for such coefficients, a (time-dependent) critical cluster size $i_c(t)$ exists, defined by
\begin{equation*}
a_iC_1(t)-b_i<0\,,\forall i<i_c\,,\quad a_iC_1(t)-b_i\geq 0\,,\forall i\geq i_c\,.
\end{equation*}
This observation has led \cite{Schweitzler} to analyze the SBD with the Ostwald ripening theory in mind. Specifically, with the help of numerical simulations, and heuristically derived moment closure approximation of the master equation \eqref{eq:mastereq} governing the clusters' distribution evolution (which resemble second-order approximation of the deterministic BD system, see Remark \ref{rmk:second_order}), the authors put in evidence the existence of a (stochastic) metastable state which is reached before the equilibrium distribution. Indeed, starting from an initial pure-monomer condition, on can observe a rapid transient that lead to a relatively small cluster distribution (with support contained among the size below the critical size), which has a long-lived state. Only after a first critical cluster is formed, the cluster size distribution is bimodal, given by a mixture of undercritical and overcritical clusters, until a single large cluster emerges from a competition between overcritical clusters, and its further growth is at the expense of the other clusters which now shrink. We have reproduced similar numerical simulations, with kinetic coefficients given by Eq.~\eqref{eq:example_coefficients}, in Fig~\ref{fig1} and~\ref{fig2}.

A key event in exiting the metastable state is thus the formation of an overcritical cluster. Such event may be analyzed with the help of the first passage time theory. It is important to note that, in agreement with classical metastability theory, the authors of this previous work noticed that the first time needed to form an  overcritical cluster was subjected to large fluctuations. We are not aware of any theoretical work on the metastability for the SBD system, but we may mention that several groups have recently investigated numerically the behavior of first passage time (or related quantities) in the SBD system (or related models) \cite{Bhatt2003}, \cite{Yvinec2012}, \cite{Yvinec2016}, \cite{Penrose2008}, \cite{Johansson2016}. In particular, it is tempting to use first passage time theory to define a stochastic analog of the so-called nucleation rate in the classical nucleation theory (see section \ref{sec:metastability}). However, we notice that the analytical form of such nucleation rate is unclear. In particular, what should be the quasi-stationary distribution, stochastic analogous to the metastable state derived in Section \ref{sec:metastability}?

Finally, let us mention the link with the study of the stochastic gelation time in Smoluchowsky's coagulation model, which has recently been the subject of active research. Let us defined, for $\alpha\leq 1$,
\begin{equation}\label{eq:fpt_smolu}
\tau_n^\alpha=\inf\{t>0\,:\, C_k^n(t)>0\,, \text{for some } k>\alpha n\}\,,
\end{equation}
where $\{C^n(t)\}$ is the rescaled stochastic process defined its generator in Eq.~\eqref{eq:sbd_rescaled}. It is known that for the stochastic smoluchowsky's coagulation model (see \cite{Jeon1998},\cite{Eibeck2001}, \cite{Fournier2004b}, \cite{Fournier2009}, \cite{Rezakhanlou2013}, \cite{Wagner2005}), and for specific coagulation kernel, the sequence of first passage time \eqref{eq:fpt_smolu} has a finite (zero or positive) limit as $n\to \infty$, and that the limit is linked to the gelation (loss of mass) in the deterministic Smoluchowsky's coagulation model. According to the longtime behavior theory for the deterministic Becker-D\"oring model, it is to be expected that for the SBD, such first passage time \eqref{eq:fpt_smolu} can only have infinite limit. However, rate of divergence and summary statistics (mean, variance) as $n\to \infty$ are important open questions.





%
%
%
\pagebreak

\pagebreak




\addcontentsline{toc}{section}{Acknowledgment}

\section*{Acknowledgment}
E.H. thanks the financial support of CAPES/IMPA Brazil during the post-doc at the Universidade Federal de Campina Grande (Paraiba).    R.Y thanks the Isaac Newton Institute for Mathematical Sciences, Cambridge, for support and hospitality during the programme Stochastic Dynamical Systems in Biology: Numerical Methods and Applications, where part of this work was undertaken.

\addcontentsline{toc}{section}{References}

\footnotesize 
\bibliographystyle{myplainnat}
\bibliography{bib-final}

\Addresses

\end{document}